\documentclass[11pt]{article}

\usepackage{amsfonts}
\usepackage{amsmath,amssymb}
\usepackage{bbm}
\usepackage{bm}
\usepackage{color}
\usepackage{slashed}
\usepackage{cite}
\usepackage[utf8]{inputenc}
\usepackage[toc,page]{appendix}
\usepackage{graphicx}
\usepackage{placeins}
\usepackage{caption}
\usepackage{subfigure}
\usepackage{booktabs}
\usepackage{comment}
\usepackage{physics}
\usepackage{float}
\usepackage{bbold}
\usepackage{dcolumn}
\DeclareGraphicsExtensions{.pdf,.png,.jpg}

\usepackage[colorlinks]{hyperref}
\AtBeginDocument{%
  \hypersetup{
    citecolor=blue,
    linkcolor=blue,   
    urlcolor=blue}}

\definecolor{airforceblue}{rgb}{0.36, 0.54, 0.66}
\definecolor{steelblue}{rgb}{0.27, 0.51, 0.71}
\definecolor{amber}{rgb}{1.0, 0.49, 0.0}
\definecolor{darkgreen}{rgb}{0.0, 0.5, 0.0}
\definecolor{amber}{rgb}{1.0, 0.49, 0.0}

\DeclareMathAlphabet{\mathpzc}{OT1}{pzc}{m}{it}

\graphicspath{{draft/Figures/}}

\allowdisplaybreaks[1]
\def\simg{{\ \lower-1.2pt\vbox{\hbox{\rlap{$>$}\lower6pt\vbox{\hbox{$\sim$}}}}\ }}
\def\siml{{\ \lower-1.2pt\vbox{\hbox{\rlap{$<$}\lower6pt\vbox{\hbox{$\sim$}}}}\ }}

\makeatletter \@addtoreset{equation}{section} \makeatother

\newcommand{\SVcomment}[1]{\emph{\color{red}SV:#1}}

\newcommand*{\diff}[1]{\text{d}#1}

\newcommand*{\der}[2]{\frac{d #1}{d #2}}

\graphicspath{{Figures/}}

\begin{document}
\flushbottom

\begin{titlepage}

\begin{centering}

\vfill

{\Large{\bf
Warm dark matter from a gravitational freeze-in \\in extra dimensions}}

\vspace{0.8cm}

A.~de~Giorgi$^{\rm a\,}$\footnote{arturo.degiorgi@uam.es}
and S.~Vogl$^{\rm b\,}$\footnote{ stefan.vogl@physik.uni-freiburg.de}

\vspace{0.8cm}

{\em $^{\rm{a}}$ Departamento de Fisica Teorica and Instituto de Fisica Teorica UAM/CSIC\\ Universidad Autonoma de Madrid, Cantoblanco, 28049, Madrid, Spain} 
\\
\vspace{0.15 cm}
{\em $^{\rm{b}}$Albert-Ludwigs-Universität Freiburg, Physikalisches Institut\\
Hermann-Herder-Str. 3, 79104 Freiburg,
Germany}
\vspace*{0.8cm}

\end{centering}
\vspace*{0.3cm}
 
\noindent

\textbf{Abstract}:  
We study the freeze-in of gravitationally interacting dark matter in extra dimensions. Focusing on a minimal dark matter candidate that only interacts with the SM via gravity in a five-dimensional model we find that a large range of dark matter and Kaluza-Klein graviton masses can lead to the observed relic density. The preferred values of the masses and the strength of the interaction make this scenario very hard to test in terrestrial experiments. However, significant parts of the parameter space lead to warm dark matter and can be tested by cosmological and  astrophysical observations.

\vfill

\end{titlepage}
\setcounter{footnote}{0}
\begin{section}{Introduction}

The origin and nature of dark matter (DM) are some of the greatest puzzles of physics today. Observations of the CMB by the Planck satellite allow for a percent-level determination of the  relic abundance $\Omega_{\hbox{\tiny DM}}h^2=0.1200 \pm 0.0012$ \cite{Planck:2018vyg} but experimental evidence for non-gravitational interactions of DM remain elusive. Given that all evidence for DM to date relies on gravity it is an interesting possibility that it is the only connection between DM and the Standard Model~(SM).  Different mechanisms for the production of particle DM with this property have been considered, see e.g. \cite{Garny:2015sjg,Garny:2017kha,Ford:1986sy,Tang:2017hvq,Ema:2018ucl}.
Unfortunately, the preferred mass range of the dark matter and the weak interaction strength make gravitationally produced particle DM very hard to test and no detectable cosmological or terrestrial signatures have been found so far. 
Therefore, it is very interesting to understand if there are alternatives that combine the conceptual simplicity of exclusively gravitational interactions with more promising observational signatures. 
In this work, we investigate one such possibility: extra dimensions that are open to gravity while matter is restricted to a 4D brane. We focus on a single extra dimension and consider two simple possibilities: Large Extra Dimensions (LED) \cite{Arkani-Hamed:1998jmv} and warped extra dimensions following a construction by Randall and Sundrum \cite{Randall:1999ee}.

Traditionally, one of the main motivations for extra-dimensional models in particle phenomenology was their ability to alleviate the hierarchy problem. If the effective scale of the Randall Sundrum model (RS) is $\mathcal{O}(\mbox{TeV})$ the observed dark matter abundance can be produced by a thermal freeze-out  but the parameter space is strongly constrained by LHC searches for KK-gravitons \cite{Rueter:2017nbk,deGiorgi:2021xvm}. 
In this work, we take an agnostic stance and do not impose any restrictions on the effective scale in the 5D theory.  We allow for significantly higher KK-graviton masses in the large extra dimensions model and consider a higher effective interaction scale in the Randall-Sundrum than conventionally preferred. This precludes a solution to the hierarchy problems in both of these models but allows for an alternative dark matter production mechanism: freeze-in \cite{Hall:2009bx}.

To allow for a successful freeze-in, the interactions of DM with the SM plasma have to be so small that the production rate is slow on cosmological time scales throughout the evolution and DM never reaches thermal equilibrium. The slow interactions allow for a continuous buildup of the relic density; the freeze-in mechanism has attracted significant attention in recent years in particular since it leads to a drastically different phenomenology compared to freeze-out, see e.g. \cite{Chu:2011be,Merle:2013wta,Blennow:2013jba,Co:2015pka,Hessler:2016kwm,Bernal:2017kxu,Belanger:2018ccd,Belanger:2018sti}.  
Gravitational freeze-in in extra-dimensional theories possesses some intriguing differences compared to 4D models. The most important one is  that the full tower of Kaluza-Klein gravitons contributes to the connection between the DM and the SM. 
This has to be taken into account in the freeze-in computation to get  a reliable result for the relic density. In addition, in a large part of the parameter space, the KK-gravitons are long-lived and redshift-like matter for an extended period of time before decaying to the DM. This implies that the mean momentum of the DM is significantly higher than the temperature of the SM plasma and one expects warm dark matter that can have a non-trivial impact on structure formation \cite{Jedamzik:2005sx,Bode:2000gq,Barkana:2001gr}. This motivates us to go beyond the computation of the total relic abundance and  study the velocity distribution of the DM. Previous work on related questions can be found in \cite{Bernal:2020fvw,Bernal:2020yqg}. These analyses include a spurious contribution to the DM yield from an unphysical high energy growth of the KK scattering cross section \cite{deGiorgi:2020qlg,deGiorgi:2021xvm} and do not  consider the effects on cosmological observable.

This paper is organized as follows: In Sec.~\ref{sec:model} we review the key aspects of the model under consideration before discussing the gravitational freeze-in in Sec.~\ref{sec:freeze-in}. We present theoretical consistency requirements and observational limits including a computation of the dark matter velocity distribution in Sec.~\ref{sec:theory_and_observations}. These results are then combined to trace out the allowed parameter space in the next section. Finally, we present our conclusions in Sec.~\ref{sec:Conclusions}.

\end{section}
\begin{section}{The Model}
\label{sec:model}
We work in the LED model \cite{Arkani-Hamed:1998jmv} and the RS model with two branes \cite{Randall:1999ee}. In the following we will briefly introduce the basic aspects of the model and some results that we require for our subsequent computations; a more in-depth introduction can be found in e.g. \cite{Rattazzi:2003ea,Raychaudhuri:2016kth,Kribs:2006mq}.
In the RS model, the 5D space-time is compactified under an $S^1/\mathbb{Z}^2$ orbifold symmetry yielding a 1D bulk bounded by two 4-dimensional (4D) branes located at $y=0$ and $y=\pi R$ where $y$ indicates the coordinate of the fifth dimension and $R$ its size.  Due to the symmetry of the model, it is convenient sometimes to work with the dimensionless angle $\varphi=y/R$. The action of the theories is given by Einstein-Hilbert gravity generalized to five dimensions
\begin{equation}
        S = S_{\text{bulk}}+S_{\text{UV}}+S_{\text{IR}} \ ,
    \end{equation}
with
    \begin{align}
    \label{eq:RS_Action}
            & S_{\text{bulk}} = \frac{M_5^3}{2}\int d^4x \int\limits_{-\pi}^\pi d\varphi \sqrt{G}(\mathcal{R}-2\Lambda_B ) \nonumber  ,\\
            & S_{\text{UV}} =\int d^4x\int\limits_{-\pi}^\pi d\varphi \sqrt{-g_{\text{UV}}}(-V_{\text{UV}}+\mathcal{L}_{\text{UV}})\delta(\varphi) \nonumber ,\\
            & S_{\text{IR}}=\int d^4x\int\limits_{-\pi}^\pi d\varphi \sqrt{-g_{\text{IR}}}(-V_{\text{IR}}+\mathcal{L}_{\text{IR}})\delta(\varphi-\pi) \ ,
    \end{align}
where $G$ is the determinant of the 5D metric, $\mathcal{R}$ the Ricci scalar and $M_5$ the 5D Planck mass.  $\Lambda_B$ denotes the vacuum energy of the bulk while $V_{\text{UV}}$ and $V_{\text{IR}}$ are the vacuum energy terms on the branes. The 5D LED model can be obtained as a limiting case of this if the size of the extra dimensions is taken to be large and the vacuum energies vanish.  $\mathcal{L}_{\text{IR}}$ and $\mathcal{L}_{\text{UV}}$ are the Lagrange densities of fields that are localized to the 4D branes while $g_{\text{IR/UV}}$ are the 4D metric on the respective branes. For the sake of concreteness, in the following, we will assume that the SM content of particles is localized on the IR-brane. The DM is taken to be a Dirac fermion with only gravitational interactions that is localized on the same brane as the SM fields.

The gravity sector particle content can be found by performing an expansion of the 5D metric around Minkowski. After fixing the gauge appropriately, the relevant degrees of freedom of the theory are given by a tensor field corresponding to a 5D graviton, $\hat{h}$, and a massless $y$-independent scalar field usually called \textit{radion}, $\hat{r}$. 
The radion is found to be massless in such models. The existence of a massless boson would imply a fifth long-ranged force stronger or comparable to gravity which would be in contrast with observations. To this purpose, we add a mass term $m_r$ by hand to the Lagrangian. We remain agnostic on the precise mechanism that generates such mass; different models have been proposed for this purpose, see e.g. \cite{Goldberger:1999uk, Luty:1999cz,Ponton:2001hq}. Typically such mechanisms generate a tower of scalar massive states of which the radion would be the lightest. Note, that the key features of the massless limit, such as the cancellations between different diagrams that reduce the energy growth of scattering amplitudes \cite{Chivukula:2020hvi,deGiorgi:2020qlg} hold in this case as well \cite{Chivukula:2022tla}.
  
  The 5D theory can be reduced to a 4D theory by the so-called Kaluza-Klein decomposition via an orthonormal basis of functions $\{\psi_n(y)\}_{n=0}^\infty$
\begin{equation}
   \begin{split}
        &\hat{h}_{\mu\nu}(x,y)=\sum\limits_{n=0}^\infty\frac{1}{\sqrt{R}}h^{(n)}_{\mu\nu}(x) \psi_n(y) \ ,\\
        &\hat{r}(x)=\frac{1}{\sqrt{R}} \psi_r \ r(x) \,.
   \end{split}
\end{equation}
After integrating out the 5th dimension,  a massless graviton corresponding to that of GR and an infinite tower of massive gravitons remain. Their masses and wave functions satisfy
\begin{equation}
     -\der{}{y}\left[A^4(y)\der{\psi_n}{y}\right]=m_n^2 A^2(y) \psi_n \,,
\end{equation}
where $A_\text{LED}(y)=1$ and $A_\text{RS}=e^{-\mu|y|}$ with $\mu \sim \sqrt{-\Lambda_B} R$ accounting for the non-zero curvature of the bulk.
In LED the KK-modes of the graviton are equally spaced, i.e. $m_n= n\, m_1$ where $m_1$ is the mass of the lightest graviton, and the interaction scale $\Lambda=M_{Pl}$. 
In RS the extra parameter allows a free choice of $\Lambda\approx M_{Pl}e^{-\mu\pi}$; the masses are not equally space but rather given by the zeros of the Bessel function of the first kind. For large KK-numbers their spacing approaches equality to good precision since $m_n = \gamma_n/\gamma_1 m_1 \approx \pi(n+1/4)/\gamma_1 \ m_1$ for large n, where $\gamma_n$ is the n-th zero of the first Bessel J-function $J_1(z)$.
In both cases, the scale $\Lambda$ and $m_1$ can be used as independent parameters of the model and they can be related to the fundamental parameters in the action, i.e. the bulk vacuum-energy and the 5D Planck mass. 
In the following, we keep $\Lambda$ as an independent variable since this is helpful for generalizing the results to both geometries. 

At leading order, the interaction of the massive gravitons and the radion with matter is given by
\begin{equation}
    \mathcal{L}_{\text{int}}=-\frac{1}{\Lambda}\sum\limits_{n=1}^\infty T^{\mu\nu}h^{(n)}_{\mu\nu}(x)+\frac{1}{\sqrt{6}\Lambda}rT \,,
\end{equation}
where $T^{\mu\nu}$ is the energy-momentum tensor and $T\equiv T^{\mu\nu}\eta_{\mu\nu}$.
If the $n$th graviton is significantly larger than all the SM particles its width into SM final states is \cite{deGiorgi:2021xvm}
\begin{align}
   \Gamma_n=\frac{73}{240 \pi} \frac{m_n^3}{\Lambda^2} \,.
\end{align}
If $m_{DM} \ll m_i$ the width of a KK graviton into fermionic DM particles is 
\begin{align}
   \Gamma(h_n\to \bar\psi \psi)=\frac{1}{160 \pi} \frac{m_n^3}{\Lambda^2} \,.
\end{align}
Neglecting the width into other KK-gravitons and the radion leads to a branching ratio into DM of  $\mathcal{B}(h_n\rightarrow \bar\psi \psi)\approx 6/293$.
The width of the radion into fermions is proportional to $m_r m_\psi^2$ if the decay is allowed. This leads to a suppression $m_\psi^2/m_n^2$ compared to the gravitons that is generically quite large. Thus the radion will contribute much less to the yield than the KK-gravitons and can be neglected.
Expressions of the widths including the full mass dependence can be found in e.g. \cite{deGiorgi:2021xvm,Han:1998sg} and are used in our numerical studies.

\end{section}
\begin{section}{Gravitational freeze-in}
\label{sec:freeze-in}
In order to allow for a non-thermal production of DM the production rate $\Gamma_{DM}$ has to be small compared to the Hubble rate throughout the evolution of the Universe, i.e. $\Gamma_{DM}/H \ll 1$ for $T \leq T_{r}$ where $T_{r}$ denotes the reheating temperature. Since the branching ratio of the KK-gravitons to DM is $\mathcal{O}(10^{-2})$ this can only be guaranteed if the production rate of the KK-gravitons $\Gamma_{KK} \lesssim H$. Therefore, we need to require that the gravitons  can at most barely reach thermal equilibrium. We expect that they will be out of equilibrium themselves in the bulk of the viable parameter space. 

Since there is no KK-graviton plasma the production of DM has to be initiated by SM states. Two contributions to the production are possible: a) annihilation of SM particles into DM with intermediate KK-gravitons and b) production of KK-gravitons in inverse decay followed by their subsequent decay. 
Note, however, that considering both processes leads to double-counting in the DM production rate since the on-shell contribution of the KK-gravitons to scattering is already taken into account by the (inverse) decays. 
This phenomenon is familiar from lepto- and baryogenesis and can be dealt with by real intermediate state subtraction, see e.g. \cite{Kolb:1979qa,Plumacher:1996kc,Giudice:2003jh}. This is not necessary here since our situation is a bit different. For center of mass energies, $E_{CM}> m_1$ the  cross-section is dominated by a series of peaks due to the different KK-gravitons in the tower going on shell and the off-shell part of the amplitude actually goes to zero between the peaks due to destructive interference. As long as the width of them is small compared to the mass gap $\Delta m$ the rate, which is obtained by integrating over $E_{CM}$ with the appropriate thermal weight factors, is completely dominated by the on-shell part and the off-shell contribution to scattering is negligible. 
Therefore, it is sufficient to consider the production from decays of the frozen-in KK gravitons that were in turn produced by inverse decays from the SM bath\footnote{ This logic fails if  $m_1 \gg T_r$ since in this regime the production of the intermediate KK-gravitons is exponentially suppressed at all temperatures. Then the production is dominated by the off-shell contribution of the KK-tower to scattering with interaction rate $\gamma \propto \frac{T^{12}}{m_1^4 \Lambda^4}$ \cite{Bernal:2020fvw}. This is suppressed by a higher power of the scale $\Lambda$ and the contribution only becomes  relevant for rather high values of $m_1/T_r$. We have checked that it does not contribute to the region of parameter space we are interested in.}.

Neglecting the backreaction the differential DM yield from the production of on-shell KK gravitons is given by
\begin{align}
\label{eq:diffyield}
    \frac{dY}{ dT}= - \sum_i \frac{\gamma_i}{H s T} \mathcal{B}(h_i\rightarrow \bar\psi \psi)
\end{align}
where $Y=n_{DM}/s$ and $s= 2\pi^2/ 45 g_{\ast s} T^3$ is the entropy density expressed in terms of the effective number of entropy degrees of freedom $g_{\ast s}$. Here $\gamma_i$ denotes the interaction rate density for the production of the ith KK graviton from inverse decays and is given by 
\begin{align}
 \gamma_i=n_{eq,i}\frac{K_1(m_i/T)}{K_2(m_i/T)}\Gamma_i =\frac{73 m_i^5 T K_1(m_i/T)}{96 \Lambda^2 \pi^3}\,,
\end{align}
where $K_1$ and $K_2$ denote the first and second modified Bessel function of the second kind and $n_{eq,i}$ is the equilibrium density of the $i$th graviton. 
In this work, we assume Boltzmann statistics for the KK-gravitons at all temperatures. 
For $T\gtrsim m_i$ quantum statistics should be used instead but this would prevent an analytic treatment while the quantitative correction is expected to be small. 
Since the right-hand side of Eq.~\ref{eq:diffyield} does not depend on $Y$ we can integrate it directly and obtain the contribution to the DM yield from KK-graviton $i$
\begin{align}
\label{eq:yield_integral}
    Y_i = \int^{T_r}_{T_0} dT\, \frac{\gamma_i}{H s T} \mathcal{B}(h_i\rightarrow \bar\psi \psi)=\int^{T_r}_{T_0} dT \, c \frac{m_i^5}{T^5} K_1(m_i/T)\,,
\end{align}
where $c=\frac{135 \sqrt{5} M_{Pl}}{256 \sqrt{g_{\ast,s}} g_\ast \Lambda^2 \pi^{13/2}}$ is a coefficient that collects all terms that do not depend (explicitly) on mass (temperature). We have neglected the contribution of non-SM finals states in the total width here. These can be trivially included and only lead to a minor shift in the numerical constant if phase space factors can be neglected.
If $m_1\gtrsim 200$ GeV the effective degrees of freedom do not depend on the temperature in the range that contributes significantly to the integral and we can pull $c$ out of the integral. Then
the temperature integral can be evaluated analytically for $m_i \ll T_r$ and gives
\begin{align}
\label{eq:Yinaive}
    Y_{i,light} \approx \frac{3 \pi c}{2} m_i\,.
\end{align}
 However, we need to sum over all KK-gravitons that contribute significantly to get the full result, i.e. $Y_{tot}=\sum_i Y_i$. Since the exponential suppression from the Bessel function only becomes fully effective for $m_i\gtrsim \mbox{a few}\times T$ we need to include modes for which $m_i \ll T_r$ does not hold. This seems to indicate that a numerical approach is needed. However, for the lightest KK-graviton  somewhat below the reheating temperature  we can approximate the KK-tower as a continuum and replace $\sum_i \rightarrow \int dn = \int \frac{1}{\Delta m} dm$ where $\Delta m$ is the mass difference between the KK-modes. For large extra dimensions $\Delta m= m_1$. In warped extra dimensions, the KK-modes are not equally spaced but the mass splitting approaches a constant for high modes. Therefore, we can use the same approximation but with $\Delta m= m_1 \pi /\gamma_1$ where  $\gamma_1$ is the first zero of the first Bessel function $J_1$.  We have checked numerically that the continuum limit leads to an excellent approximation of the sum in both cases.
 In addition, the result has a limited UV sensitivity since the Bessel function suppresses the contribution of modes with $m \gg T_r$ exponentially\footnote{We find numerically that, for a given temperature $T$, the difference between the full mass integral and one truncated at $m= 15\, T$ is $\mathcal{O}(10^{-3})$. In the following we, therefore, take $15\, T_{r}$ as an estimate of the highest scale that enters in our results.}.  
 Thus
\begin{align}
Y_{tot} \approx \frac{c}{\Delta m}\int_0^\infty dm \int_0^{T_r} dT \, \frac{m^5}{T^5} K_1(m/T)\,.
\end{align}
These integrals can be evaluated analytically and we find 
\begin{align}
  Y_{tot} \approx  \frac{45 \pi }{4} \frac{c T_r^2}{\Delta m}  \, ,
\end{align} which is one of our mains results for the freeze-in contribution.
With this expression it is possible to directly make a prediction for the dark matter relic density $\Omega h^2$ that has a simple analytic form and only depends on the four quantities $m_{DM}, \Delta m, T_{r}$ and  $\Lambda$. Since the relic density is measured to percent level precision we can use this to predict one of them in terms of the other. We find that
\begin{align}
\label{eq:TRH}
    T_{r}\approx 1.28\times 10^{-12} \sqrt{\frac{\Delta m}{ m_{DM}}} \Lambda 
\end{align}
has to be fulfilled if the freeze-in production is to account for all the dark matter in the Universe.  
Before going further it is worthwhile to note that the result seems reasonable at first sight. On the one hand, the very small numerical prefactor indicates that $T_{r}$ is much smaller than the scale of the effective theory $\Lambda$ unless the hierarchy between $m_{DM}$ and $\Delta m$, which is approximately $m_1$, is excessive. 
On the other hand, there is ample space for $\Delta m \gg m_{DM}$ which justifies neglecting $m_{DM}$ everywhere except in $\Omega_{DM}$. In particular, in the large extra dimensions scenario, a large reheating temperature is preferred since $\Lambda = M_{Pl}$ in this case.

The above result only holds if the continuum limit for the sum over the KK gravitons is justified and the lower limit of the integration is irrelevant, i.e. for $m_1 \ll T_r$. If the first KK mode has a mass of the order of the reheating temperature we should use the sum instead. This can be done numerically in this limit since only a limited number of modes contribute significantly here. However, it is also possible to find a good analytic approximation for this case.  Switching to the variable $x\equiv m/T$, we can write the integral as
\begin{equation}
   Y_i = c m_i \int\limits_{x_i}^\infty \diff{x}\ K_1(x) x^3\,,
\end{equation}
where $x_i\equiv m_i/T_r$. For $x_i> 1$ we expand the integrand for large $x$ as $K_1(x)\approx\sqrt{\frac{\pi}{2}}\frac{1}{\sqrt{x}} e^{-x}$ which leads to
\begin{equation}
\label{eq:Yiheavy}
\begin{split}
    Y_{i,heavy}&= c m_i \frac{1}{8} \sqrt{\frac{\pi }{2}} e^{-x_i} \left(15 \sqrt{\pi } e^{x_i} \text{erfc} \sqrt{x_i}+20 x_i^{3/2}+8 x_i^{5/2}+30 \sqrt{x_i}\right)\,,\\
    &\approx c m_i \frac{1}{4} \sqrt{\frac{\pi }{2}} e^{-x_i} \sqrt{x_i} \left(4 x_i^2+10 x_i+15\right)\,. 
    \end{split}
\end{equation}
The result on the first line is valid even for values of $x_i\sim 1$ with imprecision of $\mathcal{O}(10\%)$ while the second line is only correct to $\mathcal{O}(30\%)$ for such low values. Naturally, the accuracy of both results improves rapidly as $x_i$ increases. The advantage of the approximation in the second line is that  one can now  perform the sum over all gravitons analytically. Taking $m_n \approx n m_1$ we find
\begin{equation}
    Y_{tot}\approx\sum\limits_{i=1}^\infty Y_{i,heavy} = \frac{c m_1 }{4} \sqrt{\frac{\pi }{2}} \sqrt{x_1} \left(4 x_1^2 \text{Li}_{-\frac{7}{2}}\left(e^{-x_1}\right)+10 x_1 \text{Li}_{-\frac{5}{2}}\left(e^{-x_1}\right)+15 \text{Li}_{-\frac{3}{2}}\left(e^{-x_1}\right)\right)\,,
\end{equation}
where $\text{Li}_n(z)$ is the polylogarithm. For $x_1 \gg 1$ the importance of the heavier gravitons diminishes and for $x_1\gtrsim 6.8$ the lightest KK mode contributes more than $99\%$ of the total yield.  Thus we recover \ref{eq:Yiheavy} which features an overall suppression by $e^{-x_1}$ as expected for particles much heavier than the temperature. 

\end{section}

\begin{section}{Theoretical and observational limits}
\label{sec:theory_and_observations}
\begin{subsection}{Theoretical consistency}
{\bf Unitarity:}
Just like general relativity, the theory of extra-dimensional gravity is  an effective theory both in 5D and in 4D. If our results are to be trusted we need to ensure that the breakdown scale of the 4D EFT is larger than the highest center of mass-energy that matters in our computation. Conservatively, we take this to the mass of the most massive KK-mode that is relevant which we have estimated to be $m_{max}\approx 15 \,T_r$ in the previous section. As has been shown in detail, \cite{deGiorgi:2020qlg} perturbative unitarity is expected to fail at $\Lambda_{u}\approx (m_1 \Lambda^2)^{1/3}$, see also \cite{Schwartz:2003vj} for a related discussion. Thus we required $T_r \leq \frac{1}{15}(m_1 \Lambda^2)^{1/3}$.

{\bf Narrow width:}
One might be concerned that evaluating the mass integrals up to infinity leads to problems since the width of the resonance will become comparable to the mass gap at some point. In this case, the scattering amplitude cannot be modelled as a series of well-separated peaks anymore and the narrow width approximation breaks down. This invalidates our assumption that the DM production can be treated as a sum over individual resonances that are produced by inverse decays and subsequently decay to DM.  Therefore, our approximation is justified as long as the width of the resonances up to this mass is small compared to $\Delta m$.
Again we can use $m_{max}\approx 15 \,T_r$ and demand that $\Gamma_{max} \lesssim  \frac{1}{10} \Delta m$. This yields a limit on $T_r$ in terms of $\Delta m$ and $\Lambda$ that is essentially the same we got from the perturbative unitarity estimate above. 
Therefore, our approximation holds when perturbative unitarity is not violated.

{\bf Thermalization:} In addition, we need to ensure that the simplifying assumptions that enter our computation of the relic density hold. 
Concretely, we used the freeze-in approximation and thus neglected the depletion of DM particles  due to inverse decays. A simple criterion for this is that the KK-gravitons should at most barely reach equilibrium which happens for $\Gamma_i/H|_{T=m_i}\approx 1$ for modes with $m_i\leq T_r$. The DM production rate is suppressed compared to this due to the branching ratio $\mathcal{B}=\mathcal{O}(10^{-2})$ and thus the associated DM production rate is always about two orders of magnitude smaller than the Hubble rate. All lighter gravitons will fulfill $\Gamma_i/H|_{T=m_i}\approx 1$ if it holds for a graviton with $m_i=T_r$. For heavier gravitons, one should compare the width to the highest expansion rate $H|_{T_r}$ instead. However, since their production is suppressed by their mass one might wonder if restricting their width further is necessary. We have studied this numerically and find that  $\Gamma_i|_{m_i=T_r}= H|_{T_r}$ is also sufficient to avoid a significant depletion of the DM abundance in this case and adopt this criterion as our limit on thermalization.

\end{subsection}
\begin{subsection}{Cosmological bounds}

{\bf Warm dark matter:} Since the KK-gravitons are long-lived they redshift for an extended period of time as matter before they decay. The momentum of the DM particles produced by these decays significantly exceeds the mean momentum of the SM plasma at this point. Therefore, it is not guaranteed that the DM is cold and we could also face a warm dark matter scenario which is constrained by observations.

Treating non-standard models of warm dark matter correctly is a subtle problem since the precise mapping of the dark matter velocity distribution to effects on structure formation is complicated and requires detailed cosmological simulations, see e.g. \cite{Murgia:2017lwo}. These are not available in our case and performing them goes beyond the scope of this work. Therefore, we opt for a simplified approach and approximately match our scenario to the well-studied case of thermal warm dark matter (thWDM). Limits on thWDM are conventionally reported in terms of the mass $m_{\mbox{\tiny thWDM}}$ of a dark matter particle with a thermal distribution function. Using the assumed velocity distribution this can be translated into the root-mean-square velocity of the DM today \cite{Bode:2000gq,Barkana:2001gr}
\begin{align}
     v_{rms}\approx 0.04 \left(\frac{\Omega h^2}{0.12}\right)^{1/3}  \left(\frac{m_{\mbox {\tiny thWDM}}}{1\,\mbox{keV}}\right)^{-4/3}\frac{\mbox{km}}{\mbox{s}}\,.
\end{align}
Different astrophysical observables can be used to constrain $m_{\mbox{\tiny thWDM}}$. 
Typically masses larger than a few keV are required to avoid conflict with the data, 
see e.g. \cite{Irsic:2017ixq,Dekker:2021scf,Hsueh:2019ynk,Gilman:2019nap}. 
In the following, we take the limit $m_{\mbox{\tiny WDM}} \geq 3.5$ keV  \cite{Irsic:2017ixq} which is based on an analysis of  Lyman-$\alpha$ forest observations. This benchmark leads to the upper bound $v_{rms}\leq 7.5 \,\mbox{m/s}$. We will neglect the structure of the velocity distribution and apply this limit directly to the second moment of the velocity distribution $v_{rms}$ as computed in our scenario.

The velocity $v_0$ of dark matter today, that is produced by the decay of a heavy, non-relativistic KK-graviton $i$, can be estimated as \cite{Jedamzik:2005sx}
\begin{align}
\label{eq:v0single}
v_{0,i}=\frac{p_0}{m_{DM}} \approx \frac{m_{i}}{2 m_{DM}} \frac{a_d}{a_0}= \frac{m_{i}}{2 m_{DM}} \left(\frac{g_0}{g_{d}}\right)^{1/3}\frac{T_0}{T_{d}}
\end{align}
where $a$ is the scale factor, $p$ the DM momentum, and the subscript $d$ (0) signifies that the quantity is evaluated at the time when the KK-graviton decays (today). The decays are modelled as instantaneous and we determine $T_d$ by demanding $H|_{T_d}=\Gamma_i$. This leads to $v_{0,i} \propto 1/\sqrt{m_i}$. Since we do not have a single KK-graviton but receive contributions from the whole tower we have to sum over all contributing gravitons to obtain the velocity distribution $f(v)$. In our approximation, each graviton contributes a $\delta$ function at $v_{0,i}$ that has to be weighted by their fractional abundance. Thus
\begin{align}
f(v)\propto \sum_{i=1}^{\infty}\frac{Y_i}{Y_{tot}} \frac{1}{ v^2_{0,i}}\delta(v-v_{0,i})
\end{align}
and we normalize $f(v)$ by demanding $\int dv \, v^2 f(v)=1$. 
In principle, we ought to continue numerically from here
but we find that a simple analytic approximation works well.
As before we use the continuum limit to replace the sum over the KK gravitons by an integral.
To solve it analytically we approximate $Y_{i} \approx Y_{i,heavy}$ and take the first line of Eq.~\ref{eq:Yiheavy}. 
This underestimates the contribution from the light modes but gives quite accurate results for modes with a mass of the order of the reheating temperature. As the heavy modes have the largest weight and dominate the sum we expect this approximation to work well.
We find 
\begin{align}
   f(v)= \frac{16 v_{0,T_r}^4}{945 \sqrt{\pi}v^{12}}\left( e^{-\frac{v_{0,T_r}^2}{v^2}}(30 v^4 v_{0,T_r} +20 v^2 v_{0,T_r}^3 +8 v_{0,T_r}^5) + 15 \sqrt{\pi} v^5 \mbox{erfc}[v_{0,T_r}/v]  \right)
\end{align}
where $v_{0,T_r}$ is a reference velocity definied as $v_0$ evaluated at $m=T_r$ and reads
\begin{equation}
    v_{0,T_r}\approx 2.2 \frac{\Lambda T_0 }{m_{DM} \sqrt{M_{Pl} T_r}} \,.
\end{equation}
\begin{figure}
    \centering
    \includegraphics[width=0.8\textwidth]{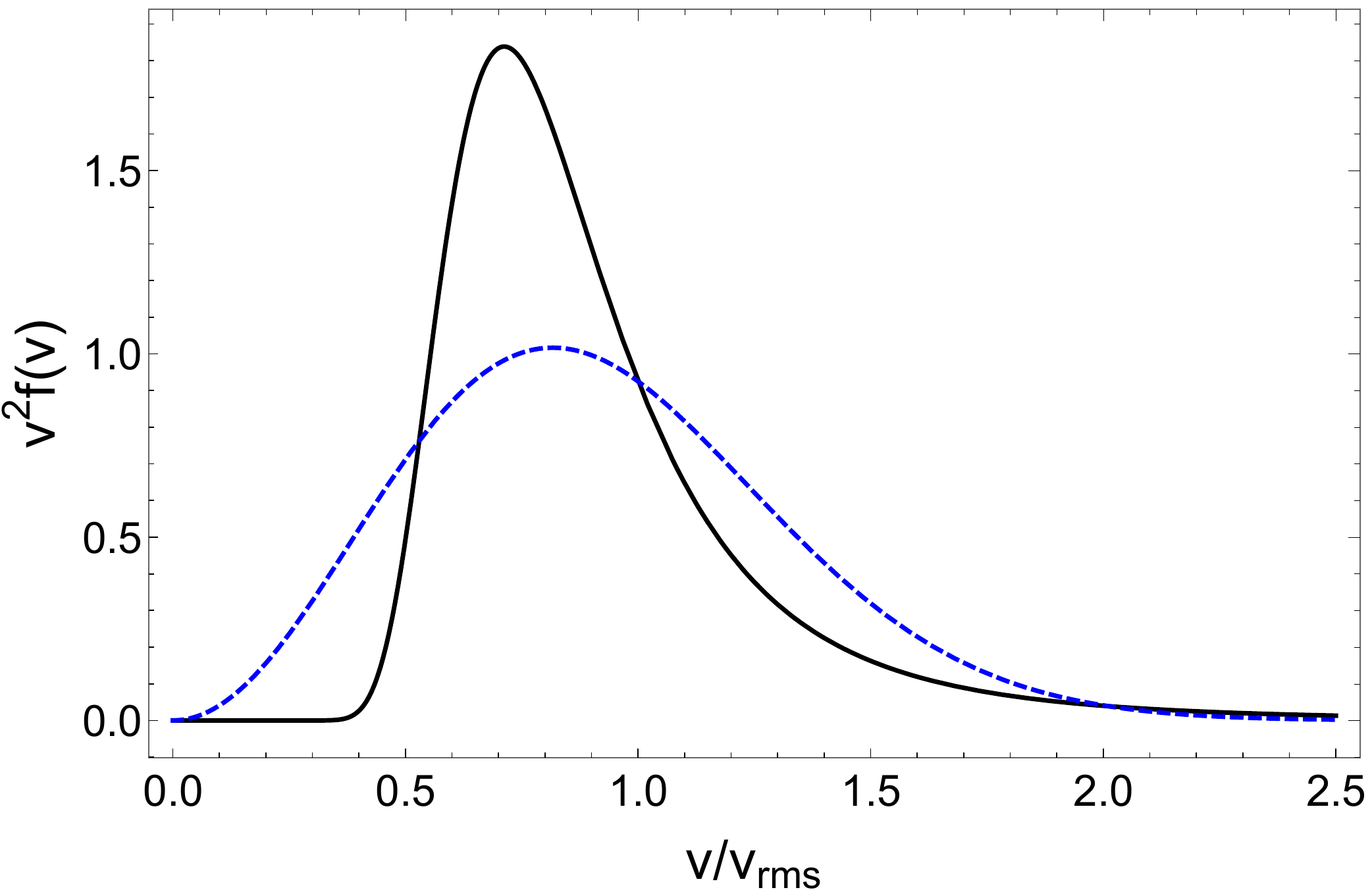}
    \caption{Velocity distribution predicted from extra-dimensional freeze-in (black, solid) as a function of velocity normalized to $v_{rms}$. For comparison, a thermal Maxwell-Boltzmann distribution with the same $v_{rms}$ is shown (blue, dashed).}
    \label{fig:velocity}
\end{figure}
A comparison of this velocity distribution with a thermal Maxwell-Boltzmann distribution is shown in Fig.~\ref{fig:velocity}. As can the seen the velocity distribution from freeze-in are qualitatively similar. However, it is more peaked than a Maxwellian and has less pronounced tails. It would be interesting to investigate which impact these differences in shape have on the astrophysical exclusions.  

By taking the second moment of the velocity distribution we find the root-mean-square velocity $v_{rms}$
\begin{align}
\label{eq:vrms}
v_{rms}^2= \int dv \, v^4 f(v)= \frac{4}{9} v_{0,T_r}^2\,.
\end{align} This result agrees with our intuition that the velocity distribution should be dominated by the KK-modes that contribute most to $Y_{tot}$, i.e. gravitons with a mass of the order of $T_r$. 
A numerical evaluation that does not use the high mass approximation for the yield leads to $v_{rms}^2\approx 0.45\, v_{0,T_r}^2$ instead. This agrees with the analytic result to about $2\%$. Expressed in terms of the parameters of our model the warm dark matter bound thus implies
\begin{align}
    \frac{\Lambda T_0 }{m_{DM} \sqrt{M_{Pl} T_r}} \lesssim 1.7 \times 10^{-8} \,.
\end{align}
 For $m_1\gtrsim T_r$ the continuum limit is not appropriate and  we have to work with the sum instead. Due to the $\delta$-function, we can perform the velocity integral straight away and get
\begin{equation}
    v_{rms}^2=\sum\limits_{i=1}^\infty\frac{Y_i}{Y_{tot}} v_{0,i}^2 \,.
    \label{eq:vrmsheavygeneral}
\end{equation}
Approximating $Y_i$ with the second line of Eq.~\ref{eq:Yiheavy} leads to
\begin{equation}
\label{eq:vrmsheavy}
\left(v^{heavy}_{rms}\right)^2\approx \frac{1}{x_1}\frac{4 x_1^2 \text{Li}_{-\frac{5}{2}}\left(e^{-x_1}\right)+10 x_1 \text{Li}_{-\frac{3}{2}}\left(e^{-x_1}\right)+15 \text{Li}_{-\frac{1}{2}}\left(e^{-x_1}\right)}{4 x_1^2 \text{Li}_{-\frac{7}{2}}\left(e^{-x_1}\right)+10 x_1  \text{Li}_{-\frac{5}{2}}\left(e^{-x_1}\right)+15  \text{Li}_{-\frac{3}{2}}\left(e^{-x_1}\right)}v_{0,T_r}^2\,.
\end{equation}
For $x_1 \rightarrow 0$ the prefactor reduces to $2/5$ which is within $10\%$ of the one obtained using the continuum integral\footnote{The difference is due to the treatment of $Y_{i,heavy}$. If we use the less accurate approximation in the second line of Eq.~\ref{eq:Yiheavy} in the continuum limit as well the results agree exactly for $x_1 \rightarrow 0$.}. For $x_1$ larger than 1, the prefactor decreases but remains within one order of magnitude up to $x_1\approx 20$. In our study of the allowed parameter space, we switch from Eq.~\ref{eq:vrms}
 to Eq.~\ref{eq:vrmsheavy} at $x_1=1$.
 
The above computation  has to be modified for modes with $\Gamma> H|_{T_r}$ since the criterion for the determination of the decay temperature we use leads to a fictional decay temperature above the reheating temperature or, equivalently, an unphysical scale factor $a_d < a_{T_r}$ in this case. Due to the thermalization bound $\Gamma_{m_i=T_{r}}\leq H|_{T_r}$ this situation can only arise at the edge of the parameter space we consider and we do not analyze it in detail. Nevertheless, it is interesting to get an estimate for the warm dark matter bound in this limit. Since most of the yield is due to modes with $\Gamma> H|_{T_r}$ in this case we can model their production and decay as instantaneous and use $T_{r}$ to determine the redshift factor that enters in eq.~\ref{eq:v0single}. The rest of the analysis goes through as before. For $m_1 \lesssim 5 T_r$
most model parameters drops out and we find $v_{rms}\approx 10^{-3}/m_{DM}[\mbox{keV]}$\footnote{For larger values of $m_1$ the other model parameters have an impact and there is an approximately linear increase of the velocity with $m_1/T_r$.}.
Thus $m_{DM} \gtrsim 6$ keV is necessary to avoid the warm dark matter bound in this case. \\

\noindent{\bf Big Bang Nucleosynthesis:}
As discussed previously the KK-gravitons are expected to be rather long-lived in the early Universe. If their lifetime $\tau$ exceeds about ten seconds they decay during Big Bang Nucleosynthesis (BBN) and can influence the production of the primordial nuclei, see e.g. \cite{Reno:1987qw,Cyburt:2002uv,Jedamzik:2006xz,Kawasaki:2004yh,Kawasaki:2017bqm}. The impact of the decays depends on the amount of energy that is injected into the SM plasma, the time at which the decays happen and the decay mode. 
The limits on the different hadronic channels are very similar and much stronger than those on leptonic ones. As KK-gravitons have an $\mathcal{O}(1)$ branching ratio to hadronic final states we consider the limits on hadronic decays and neglect the energy that is lost to other channels. Very roughly speaking the limits are strong for $\tau \geq 10 s$
 \cite{Kawasaki:2017bqm}. 
Thus, we can obtain a conservative limit by demanding that all KK-gravitons have a lifetime $\tau = \hbar / \Gamma \lesssim 10$ s. Plugging in the expression for the width leads to  $m_1/\mbox{GeV} \gtrsim 8.7 \times 10^{-9} (\Lambda/\mbox{GeV})^{2/3}$. For the LED models this  implies $m_1 \gtrsim 46$ TeV while the value is lower in the RS model. 
One might wonder if the yield of the KK gravitons is large enough to have an impact on BBN.
For the RS model, i.e. $\Lambda < M_{Pl}$, we find that the ratio of the energy density to entropy density $m_1 Y_1$ of a KK graviton with a lifetime of 10 seconds is well in excess of $10^{-14}$~GeV, which is the rough value a relic with dominantly hadronic decays needs to have in order to have an impact on BBN~\cite{Kawasaki:2017bqm}. For LED $m_1 Y_1$ approaches $10^{-14}$~GeV but is still a factor of a few larger. Therefore, we adopt $\tau_1 \leq 10$~s  as our bound in the following. 
\end{subsection}
\end{section}
\begin{section}{Results}
We now put all our previous results together.
In general, we have four independent parameters. 
Three of these come from the particle physics model and can be chosen to be $m_{DM}$,  $\Lambda$ and $m_1$\footnote{Recall that the LED model with $\Lambda=M_{Pl}$ and the Randall Sundrum model with $\Lambda< M_{Pl}$ have a different relation between $m_1$ and $m_n$.}, while the fourth is the reheating temperature which is set by the cosmological history of our Universe.
$\Omega_{DM}h^2$ is known to percent-level precision and, therefore, we can remove one of these parameters in favour of the others by demanding that the DM abundance produced by freeze-in matches the observations. 
We decided to fix the reheating temperature and keep the particle physics parameters free.  
\label{sec:Results}
\begin{figure}[htbp] 
\centering%
\subfigure[{}\label{fig:Lambda1}]%
{\includegraphics[width=0.47\textwidth]{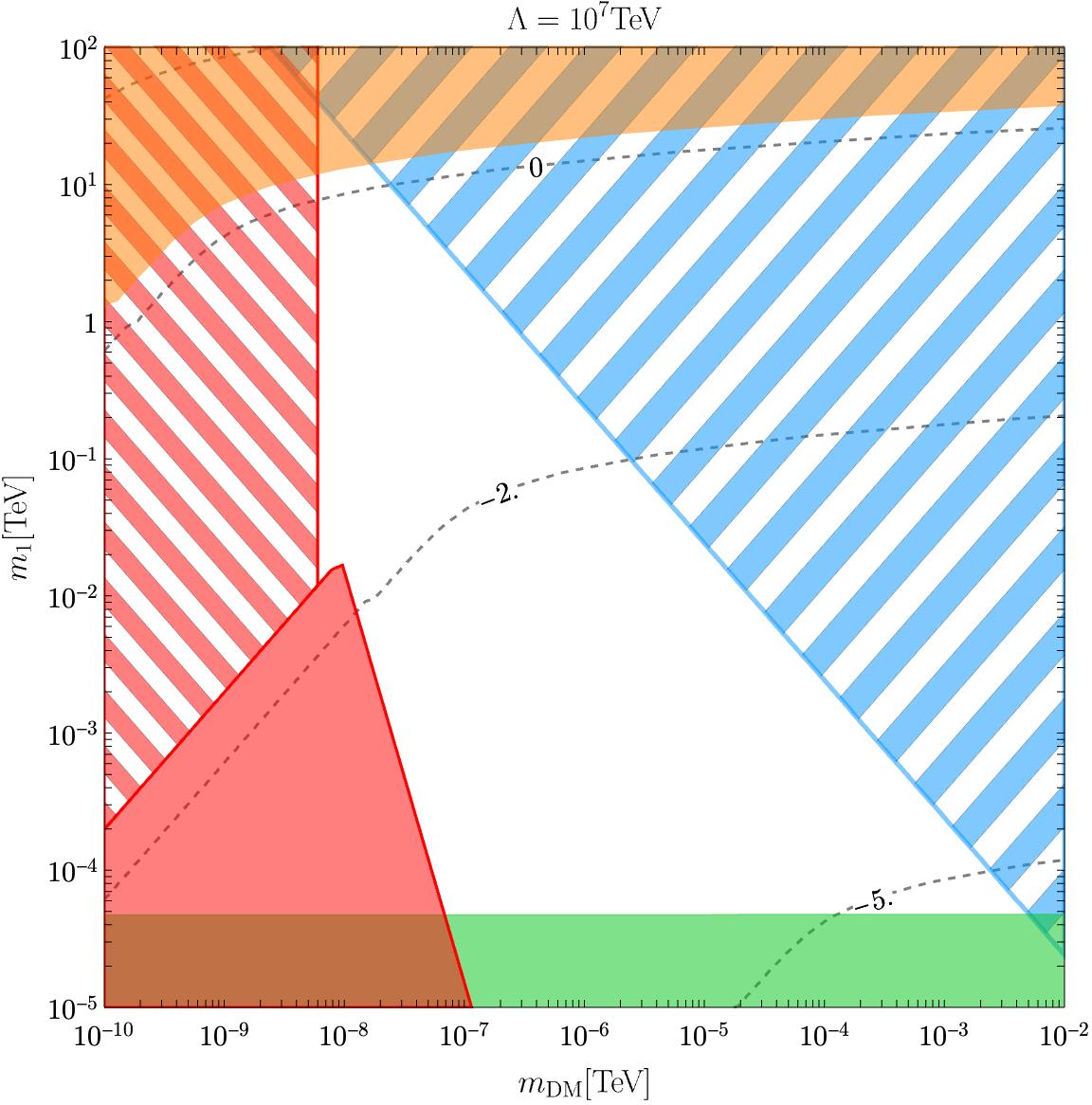}}
\subfigure[{}\label{fig:Lambda6}]%
{\includegraphics[width=0.46\textwidth]{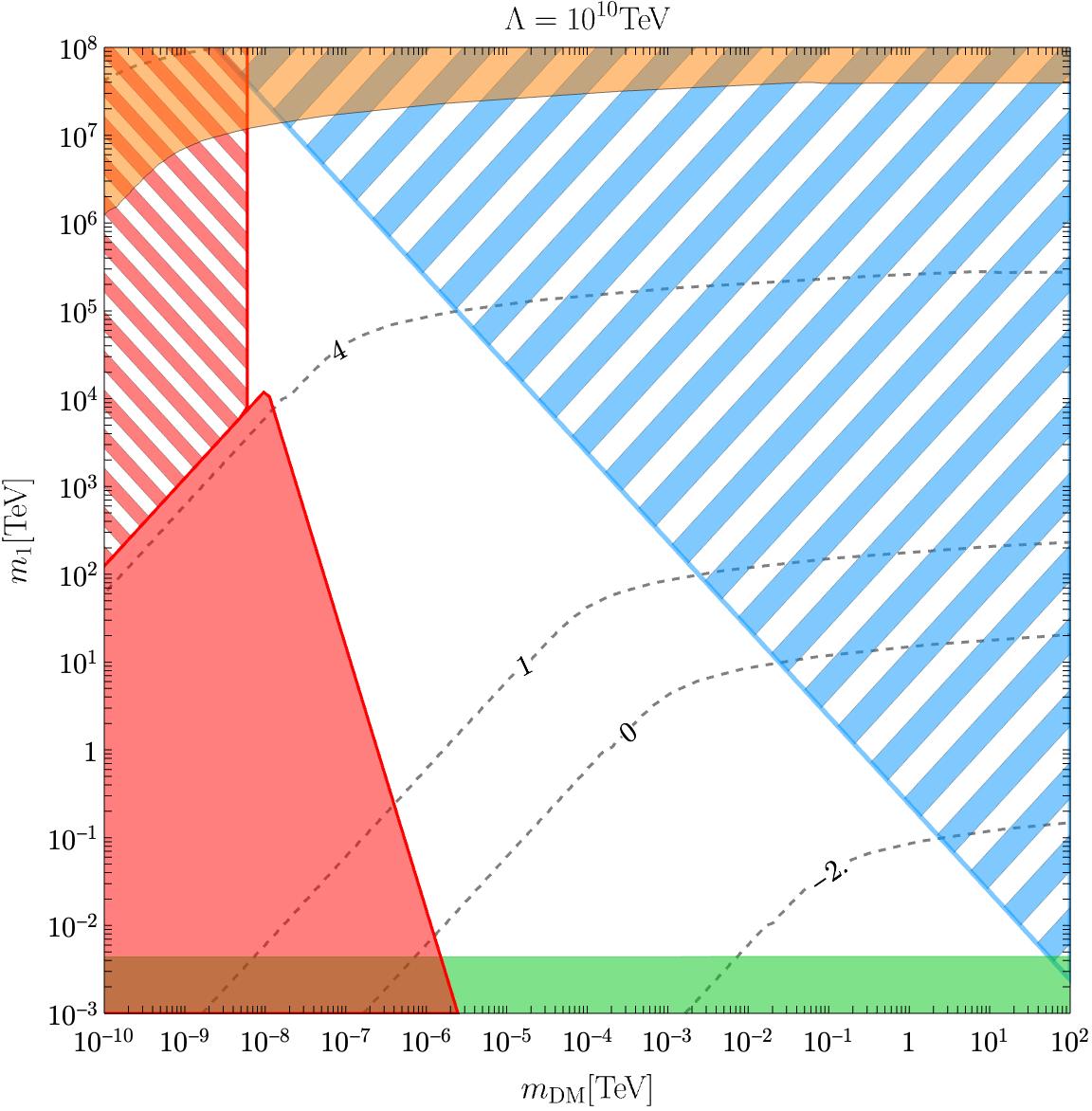}}
\subfigure[{}\label{fig:Lambda11}]%
{\includegraphics[width=0.47\textwidth]{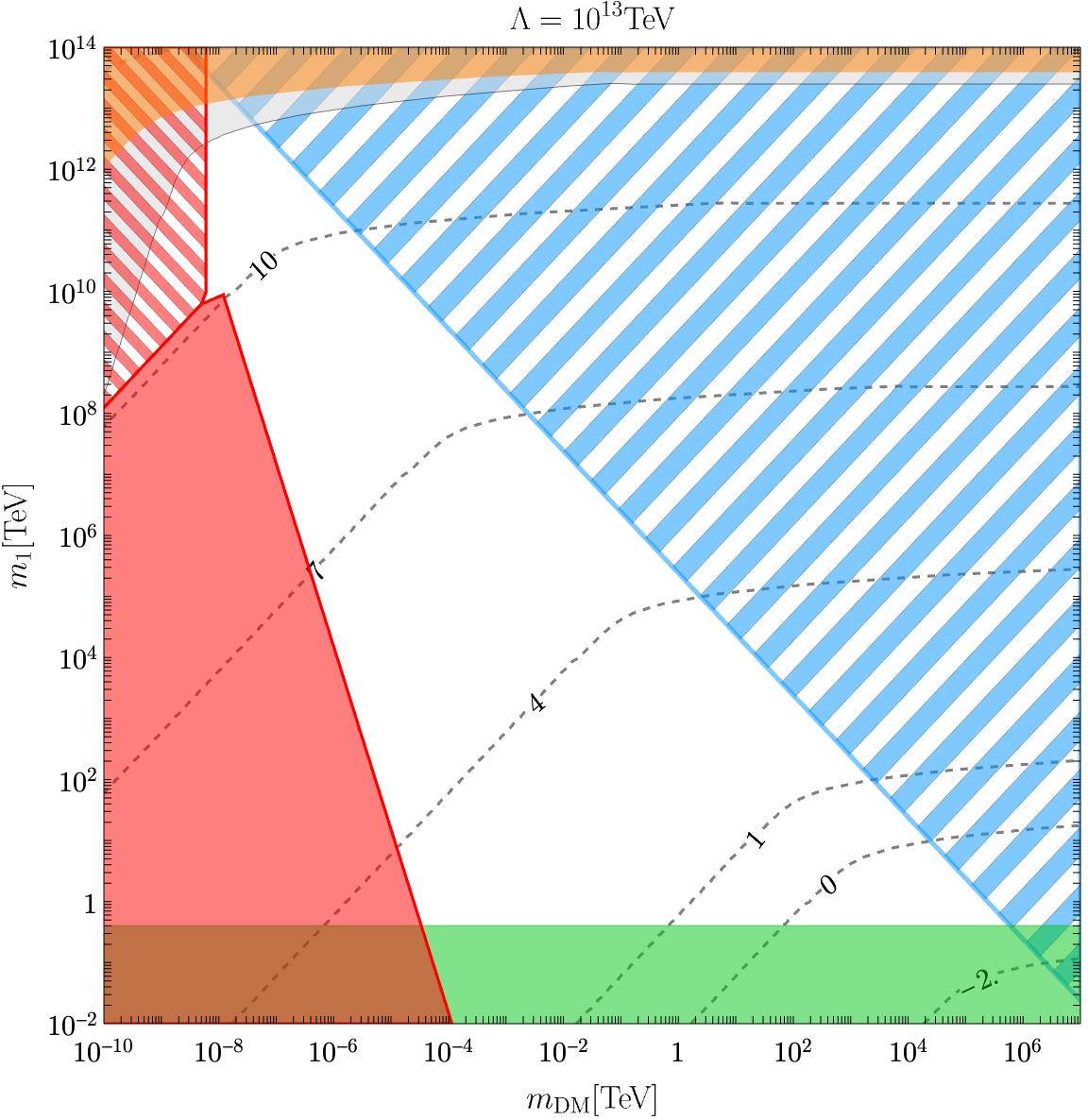}}
\subfigure[{}\label{fig:Lambda16}]%
{\includegraphics[width=0.47\textwidth]{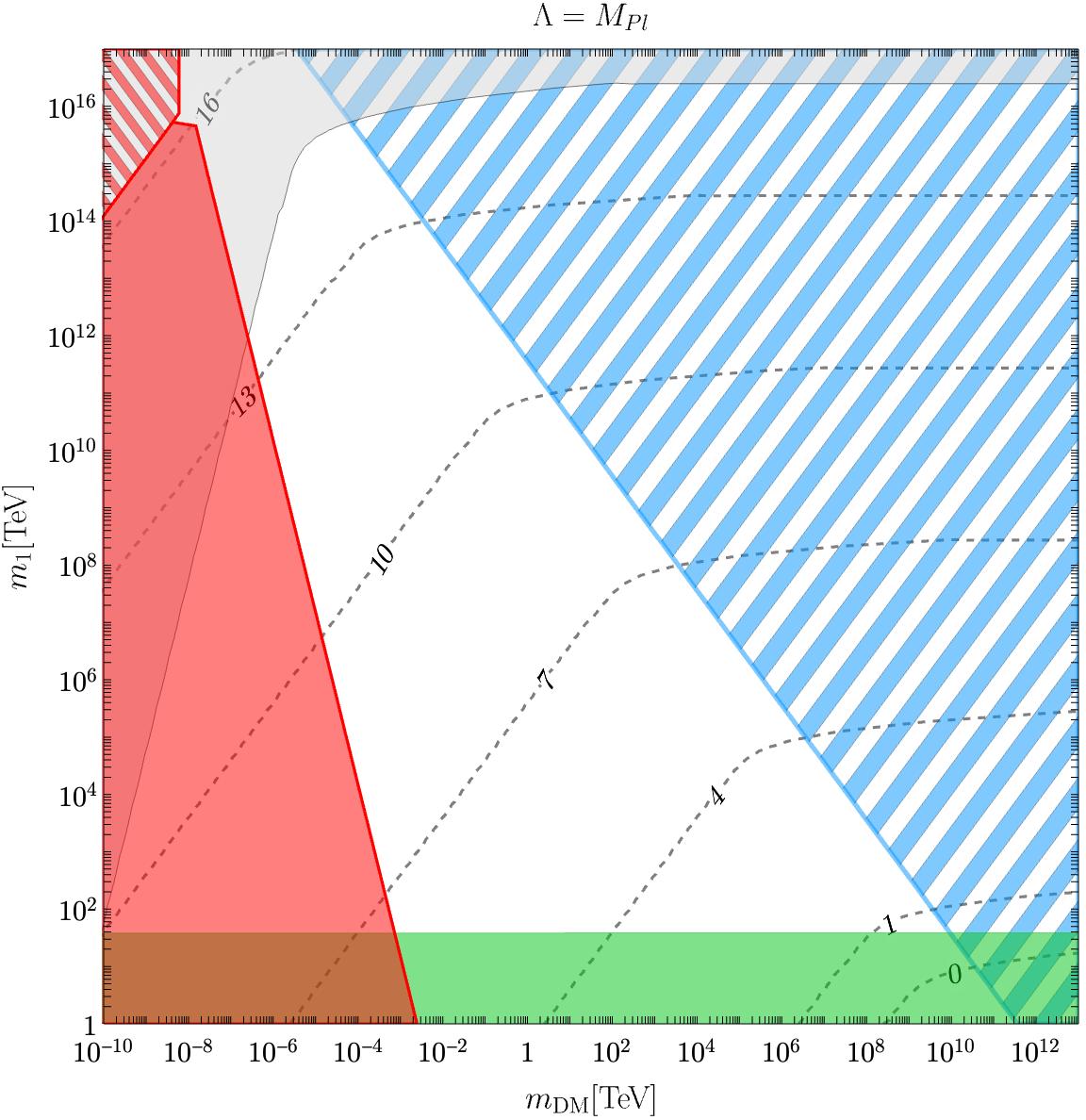}}
\caption{The coloured regions are excluded due to thermalization (orange), structure formation (red), BBN (green) and unitarity (gray). The low-reliability region of the yield and structure formation is shown in light-blue and with a meshed shading, respectively. The dashed black lines correspond to contours with constant $\log_{10} T_r$ [TeV].
\label{fig:plots} 
}
\end{figure}

In Fig.~\ref{fig:plots} we show the $m_{DM}, m_{1}$ parameter space for four representative choices of $\Lambda$. 
The parameter region that allows for successful production of the total relic density is shown in white whereas the coloured regions are in conflict with observational bounds or the consistency conditions of our calculation.
Starting from the top, the orange region is excluded as it would imply thermalization of the gravitons.
The unitarity bounds (grey) show up in a similar region and can be weaker or stronger than the thermalization limit depending on the scale $\Lambda$.
The red area on the left is excluded by the warm DM bounds. 
As we approach the thermalization limit the assumption that all KK gravitons are long-lived starts to fail and this result needs to be corrected. 
To illustrate this we cut the warm dark matter bound at $\Gamma_{m=Tr}=10^{-2}H|_{T_R}$ and show the estimate for the bound on the boundary in a meshed shading of the same color. We expect that the true result here will be a smooth transition between the two.
At the bottom, there is a bound from  BBN ($\tau_1 \leq 10$~s). 
Finally, the meshed blue area on the right shows the region where the relic density becomes very sensitive to the reheating temperature which we define as a $\gtrsim 100\%$ variation in the DM yield for a $10\%$ increase of $T_r$. Since details of the reheating process could have a significant impact on the relic density in this region the results are not reliable here and we do not consider them further. In this region, the gravitons are heavier than $T_r$ and hence it is characterized by an exponential dependence of the yield on $T_r$. For lower scales, this is already excluded by the thermalization limit.
All considered, there is ample open parameter space for $\Lambda \geq 10^7$~TeV up to $\Lambda = M_{Pl}$. A significant part of this is excluded by the warm dark matter bound and BBN. New astrophysical and cosmological observations are expected to strengthen the bounds. However, given the huge range of allowed parameters testing this mechanism fully seems out of reach unless novel observations come into play that could, for instance, allow the determination of the reheating temperature.

\end{section}
\begin{section}{Conclusions}
\label{sec:Conclusions}

In this work, we investigated  extra dimensions that are open to gravity while the matter content of the SM and the DM are restricted to a 4D brane. We focused on a single extra dimension and consider two simple possibilities: Large Extra Dimensions (LED) and warped extra dimensions in the Randall Sundrum (RS) model.
A notable difference compared with gravitational interactions in 4D is the presence of a tower of Kaluza-Klein gravitons that provide a new connection between the DM and the SM.
If the 5D Planck mass is sufficiently high compared to the reheating temperature of the universe the KK-gravitons do not reach thermal equilibrium and are produced by a freeze-in process. 
Their subsequent decays to DM lead to a cosmological slow buildup of DM abundance. 
We sum the contribution from the whole tower and find simple analytic expressions for the expected DM yield. 
We find that a large range of DM and KK graviton masses allows accommodating the observed relic density. 
Due to the relatively high scales preferred by the relic density, testing this idea with terrestrial experiments seems very challenging. 
However, there are interesting consequences for cosmology that can be used to constrain the allowed parameter space. 
The most important one is the prediction of warm dark matter. 
Since the KK gravitons are comparatively long-lived they redshift for an extended period of time before they decay and produce DM particles with a momentum that significantly  exceeds the characteristic momentum of the SM plasma. 
The ensuing velocity distribution of the DM is non-thermal and does not follow a Maxwell-Boltzmann distribution at late times.  
It would be interesting to investigate how the non-thermal shape of the distribution affects the structure formation process but a detailed study of this goes beyond the scope of this work. 
Therefore, we employed a simple estimate to translate the limits on thermal warm dark matter in the literature to our scenario by comparing the root mean square velocity predicted from our model to the limit for thermal warm dark matter. We find that DM masses up to $\mathcal{O}(1)$ GeV can be excluded by existing bounds. The search for the impact of non-cold DM on the formation of structure at small scales is expected to make significant progress in the next years and, therefore, further parts of the parameter space will be probed soon. However, a complete test of the allowed parameter space seems very challenging at present.

\end{section}

\section*{Acknowledgments} 
This project has received funding /support from the European Union’s Horizon 2020 research and innovation programme under the Marie Skłodowska-Curie grant agreement No~860881-HIDDeN.

\begin{appendices}
\numberwithin{equation}{section}


\end{appendices}

\bibliographystyle{hieeetr}
\bibliography{extra_dim_freeze-in.bib}

\end{document}